# Integrated GaN photonic circuits on silicon (100) for second harmonic generation

Chi Xiong[1,†], Wolfram Pernice[1,†], Kevin K. Ryu[2,†], Carsten Schuck[1], King Y. Fong[1], Tomas Palacios[2] and Hong X. Tang[1,*]

[1]*Department of Electrical Engineering, Yale University, New Haven, CT 06511, USA*
[2]*Department of Electrical Engineering and Computer Science, Massachusetts Institute of Technology, Cambridge, MA 02139, USA*
[†]*These authors contributed equally to this work*
[*]*hong.tang@yale.edu*

**Abstract:** We demonstrate second order optical nonlinearity in a silicon architecture through heterogeneous integration of single-crystalline gallium nitride (GaN) on silicon (100) substrates. By engineering GaN microrings for dual resonance around 1560 nm and 780 nm, we achieve efficient, tunable second harmonic generation at 780 nm. The $\chi^{(2)}$ nonlinear susceptibility is measured to be as high as $16 \pm 7$ pm/V. Because GaN has a wideband transparency window covering ultraviolet, visible and infrared wavelengths, our platform provides a viable route for the on-chip generation of optical wavelengths in both the far infrared and near-UV through a combination of $\chi^{(2)}$ enabled sum-/difference-frequency processes.

## 1. Introduction

Second order optical nonlinearity ($\chi^{(2)}$) in crystals such as lithium niobate (LiNbO$_3$) and potassium titanyl phosphate (KTP) has been widely exploited in modern optics for frequency doubling [1], parametric down conversion [2] and sum-/difference- frequency generation [3]. While optical nonlinearity is of importance for classical non-linear optics it is also an essential resource for the generation of non-classical light and applications in quantum optics [4, 5]. Furthermore, $\chi^{(2)}$ nonlinearity is the prerequisite for a number of key technological components such as electro-optical modulators [6, 7] and optical parametric oscillators [8].

Nonlinear experiments typically rely on precise interplay between a large number of linear and non-linear components. In order to move beyond table-top setups, chip scale integration is highly desirable. In this context silicon photonics has emerged as a promising platform for arranging optical devices in a scalable fashion. Silicon's centrosymmetric lattice structure, however, does not permit second-order nonlinearity. An approach to overcome this limitation is to integrate materials with large intrinsic $\chi^{(2)}$ nonlinearity into silicon photonics. This heterogeneous integration of different material systems has been successful in delivering complementary optical characteristics to a monolithic platform, for example in the demonstration of the hybrid silicon laser [9] and germanium avalanche photodetectors [10].

Here we present a new technique to heterogeneously integrate gallium nitride (GaN) on CMOS compatible substrates as a linear and non-linear optical material for integrated photonic devices. GaN exhibits strong second-order nonlinearity [11-14], with a $\chi^{(2)}$ coefficient of the same order as LiNbO$_3$. Its outstanding electrical, thermal and optoelectronic properties have enabled a broad range of technological applications including high speed and high power electronics [15], blue/UV light emitting and laser diodes [16]. Furthermore, GaN has an extremely wide transparency window spanning UV, visible and far infrared wavelengths. Employing GaN thin films on silicon (100) substrates, we fabricate nanophotonic circuitry and we demonstrate strong second harmonic generation (SHG) on chip.

## 2. Fabrication of GaN-on-insulator substrates

Waveguiding in GaN (refractive index n ~ 2.3 at 1.55 μm) requires cladding layers of lower refractive index to confine light to the GaN layer. In silicon photonics, this requirement is fulfilled by utilizing silicon thin films on top of a low-refractive index silicon dioxide buffer layer, thermally grown on bare silicon substrates. The CMOS compatibility of such silicon-on-insulator (SOI) substrates is one reason for the success of silicon photonics. However, integration of GaN on a CMOS compatible substrate is challenging. Growth of high quality GaN by metalorganic chemical vapor deposition (MOCVD) is normally viable only on closely lattice matched substrates such as sapphire, silicon carbide and silicon (111). Techniques to form GaN on a silicon substrate, most notably epitaxial growth and lift-off [17, 18], encounter difficulties when separating the grown films from the growth substrate because of the strong bonding at the interface.

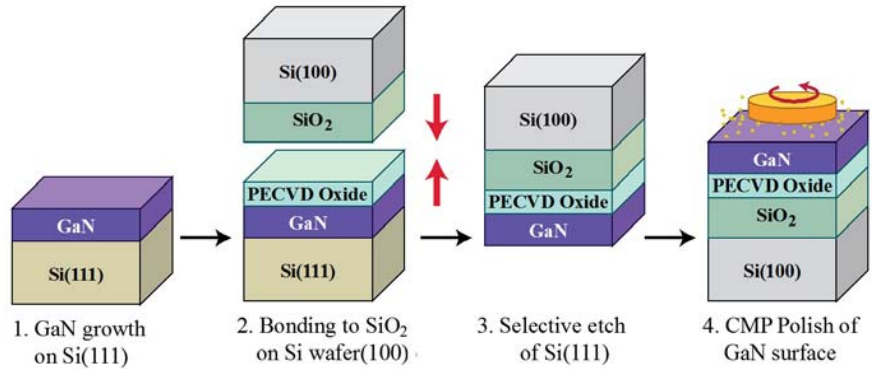

Fig. 1. GaN photonic circuits are built on GaN on silicon dioxide on silicon (GaNOI) substrates. The GaNOI substrate fabrication process runs as follows: PECVD oxide is first deposited onto commercially available GaN on Si (111) wafers to assist the bonding process; the Si (111) layer is removed after the bonding and the new GaN surface undergoes CMP.

To overcome these limitations, we developed a robust bonding process that allows us to realize photonic structures in GaN thin films atop silicon dioxide on silicon substrates (GaNOI). As illustrated in Fig. 1, we employ commercially available crystalline GaN (1.8 μm thickness) on silicon (111) wafers from Nitronex Corporation. A layer of silicon dioxide with a thickness of 1.5 μm is deposited onto the GaN films by plasma-enhanced chemical vapor deposition (PECVD) to assist the bonding process [19]. The three-layer structure is bonded to silicon wafers of (100) crystal orientation, covered with a thermally grown oxide layer of 1.8 μm thickness. The bonding process forms a stable substrate with very minor mismatch and residual strain. Subsequently the silicon (111) bottom layer is removed using sulfur hexafluoride ($SF_6$) deep reactive ion etching (RIE). In order to obtain a smooth surface, the now free GaN surface undergoes a chemical-mechanical polishing (CMP) step, reducing the GaN layer thickness to a final thickness of 400 nm. The resulting GaNOI substrates feature a root-mean-square (RMS) roughness of 4 nm as shown in the atomic force microscope (AFM) scan in Fig. 2(a).

Photonic waveguides are realized in the GaN top layer by electron beam lithography and subsequent dry etching in inductively coupled chlorine plasma. In Fig. 2(b) we show a cross-sectional scanning electron microscope (SEM) image of a cleaved fabricated device to illustrate the GaN strip waveguide structure residing on top of the bonded oxide layer. The underlying silicon (100) substrate is also discernable in the SEM picture.

## 3. Design of the GaN photonic circuits for efficient SHG

We fabricated nano-photonic circuitry on the GaNOI substrates as described in the previous section. An optical micrograph of a typical circuit is shown in Fig. 3(a). Here a microring resonator with a radius of 40 μm is coupled to the input waveguide of 860 nm width, which is separated by a gap of 150 nm from the ring resonator. Such microrings are near critical coupled at wavelengths around 1560 nm, showing extinction ratio of ~20 dB and typical measured quality factors around 10,000.

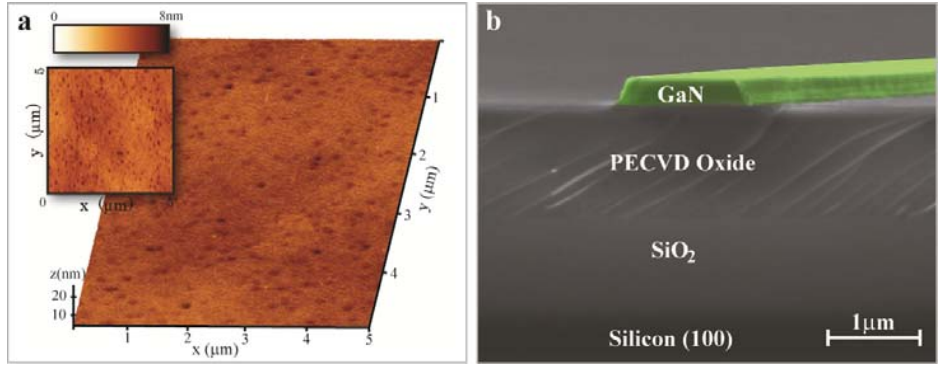

Fig. 2. a) An AFM image of a 5 μm × 5 μm area showing rms roughness of 4 nm on the polished GaN surface; (b) a cross sectional scanning electron micrograph showing a fabricated GaN waveguide (in false-color), oxide layers and the silicon (100) substrate. The trapezoidal profile of the waveguide is due to the anisotropy of the RIE process.

In order to be able to couple light into the device, we utilize focusing grating couplers which have been successfully used in silicon photonics [20]. Two sets of couplers are optimized for operation at 1560 nm and the frequency doubled regime around 780 nm by adjusting the grating period accordingly. Typical insertion loss for grating couplers at 1550 nm is measured to be 8.5±1 dB, while the couplers at 780 nm show slightly better coupling efficiency of 7.5±1 dB.

The optical response of the photonic circuits is obtained by using the experimental setup shown in Fig. 3(b). Light from a tunable IR laser source (New Focus 6428) is amplified with an erbium doped fiber amplifier (EDFA). The output power after the EDFA is monitored with an IR photodetector (New Focus 2011) and a calibrated 99:1 splitter. The amplified light is coupled into the photonic circuit using the two outer grating couplers shown in Fig. 3(a). A spectrometer (JAZ spectrometer, ocean optics) insensitive to wavelengths above 850 nm and a visible light power meter are used to analyze the light collected by the grating couplers operating at 780 nm. The 780 nm grating couplers also serve to provide further suppression of spurious pump-light by more than 40 dB. The optical response of the device is obtained by aligning single-mode 1550 nm fibers to the input and output couplers and measuring the power spectrum with a tunable laser source. Due to the unavailability of a wide-band tunable laser source in the near-IR wavelengths, the optical response of the microring around 780 nm is characterized by launching a broad-band light from a superluminescent diode (QPhotonics QSDM-790-2) and recording the transmission spectrum on a optical spectrum analyzer (HP70952B). Limited by the resolution of the OSA, we measured the loaded NIR quality factor to be $3,000 \pm 1,500$.

While critical coupling at 1560 nm requires a relatively large gap of 150 nm, the situation is different for the frequency doubled regime at 780 nm. As pointed out in reference [21], critical coupling at visible wavelengths requires significantly reduced coupling gaps below 100 nm because the waveguide mode at 780 nm extends less beyond the waveguide boundary. Introducing a drop port to the ring resonator at such short distance will lead to severe over-coupling at 1560 nm and thus compromise the optical quality factor. Therefore we employ the "pulley" waveguide structure [21] shown enlarged in the inset of Fig. 3(a) to enhance the interaction length with the ring resonator and selectively couple the generated SHG out of the microring.

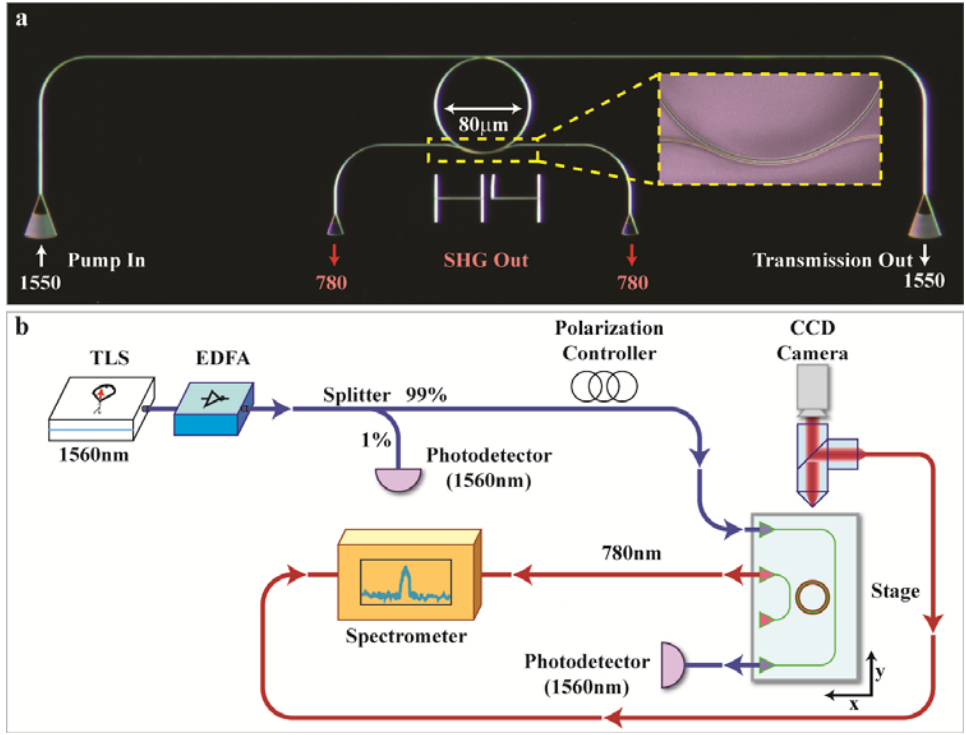

Fig. 3. Photonic device and measurement setup. (a) optical micrograph of the GaN ring resonator circuit; the pump light is coupled through one of the 1550 nm grating couplers (outer pair). The generated SH is coupled out from the center 780 nm grating coupler; the inset shows a false color scanning electron micrograph of the coupling region showing the 500 nm wide pulley waveguide coupled to the ring resonator with a coupling gap of 150 nm. The waveguide width of the ring is 860 nm. (b) measurement setup: the microring is pumped optically by a tunable laser source (TLS), amplified by an erbium doped fiber amplifier (EDFA); the SH generated light is coupled out of the device into optical fibers and analyzed with a spectrometer. Meanwhile the out of plane scattered SH light is collected by a microscope objective and focused onto a multimode fiber with a 50 μm diameter core. A visible spectrometer is used to analyze the wavelength spectrum.

Compared with one point coupling, the pulley waveguide is designed to have a 24.5 degree coupling angle with respect to the ring resonator, allowing for a larger coupling gap and hence easier fabrication. The coupling angle corresponds to a coupling length of 17.5 μm to provide maximum power transfer from the ring into the drop waveguide at an out-coupling gap of 150 nm. The waveguide width of the output waveguide is optimized by finite-difference time-domain (FDTD) simulations to provide optimal out-coupling characteristics between the fundamental mode at 1560 nm and the second harmonic (SH) at 780 nm. The resulting reduced width of the drop waveguide of 500 nm is in the vicinity of the cut-off width for the pump light at 1560 nm, thus preventing transfer of the pump light into the drop port.

### 4. Direct imaging and measurement of second harmonic generation

We employ the device shown in Fig. 3(a) for nonlinear operation. Due to the intrinsic second order non-linearity of GaN, frequency doubled light can be generated when phase matching between the pump wavelength and the second harmonic component is achieved. Because the grating couplers are matched to TE-polarized light and the waveguides at 1560 nm only support the fundamental quasi-TE mode, we approximate GaN as an isotropic material in-plane. The chromatic dispersion in nanophotonic waveguides can be compensated by varying the geometry of the waveguide, thus inducing geometric waveguide dispersion of opposite sign [22]. This is illustrated with finite-element (FEM) simulations in Fig. 4(a-c).

The waveguide side-walls are tilted from the vertical direction by an angle of roughly 8 degrees, which is extracted from the SEM image of fabricated devices as shown in Fig. 2(b). Keeping the waveguide height constant at 400 nm we vary the waveguide width which alters the amount of geometric dispersion for a given wavelength. Phase matching is achieved when the geometric dispersion exactly compensates the chromatic dispersion at the zero-dispersion width. In Fig. 4(a) we illustrate this scenario for phase-matching between the fundamental mode at 1560 nm, shown in Fig. 4(b), and the 6th order mode at 780 nm, shown in Fig. 4(c). For a waveguide width of 860 nm both waveguide modes feature identical effective refractive indices, indicating that the phase-matching conditions [23, 24] for second harmonic generation are fulfilled.

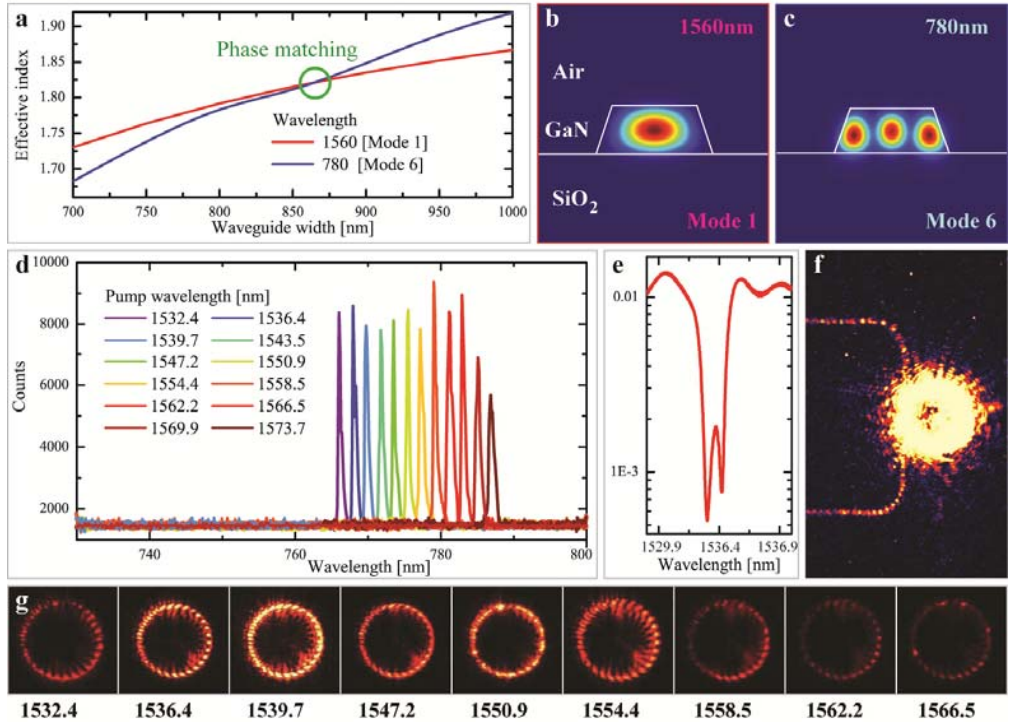

Fig. 4. Second harmonic generation in a GaN microring resonator. (a) phase matching is achieved by varying the waveguide width to match the effective index of the fundamental pump light to that of the sixth optical mode of the SH. The optimal waveguide width is found to be 860 nm. (b) FEM simulation of fundamental mode profile for 1560 nm wavelength at the optimal waveguide width of 860 nm; the waveguide sidewalls are non-vertical due to chemical RIE. (c) FEM simulation of the sixth mode profile of the SH wavelength of 780 nm at the optimal waveguide width of 860 nm.(d) tuning of SHG from 766 nm to 788 nm by aligning the pump light to different resonant wavelengths; (e) the double dips in the optical resonances confirm that both clockwise and counterclockwise modes are excited in the ring resonator; (f) visible CCD camera image (in false-color) showing SH is coupled out of the ring resonator in both directions of the output waveguide. The picture is taken with an exposure time 60 times longer than that used in panel (g); (g) a series of visible CCD camera images (in false color) showing SHG standing wave patterns on the microring when the pump wavelength is tuned into different resonant wavelengths.

The calculation takes into account the trapezoidal form of the waveguides which are a result of the dry etching procedure. The side-wall angle is obtained from the SEM image in Fig.2b) and estimated to be roughly 8 degrees.

We demonstrate second harmonic generation experimentally using the setup described in Fig. 3(b). When the input laser wavelength is swept continuously from 1532 nm to 1574 nm we observe several optical resonances in the IR spectrum, corresponding to the wavelengths at which the microring resonance condition is fulfilled, separated by the free-spectral range (FSR) of the ring which is 3.8 nm around 1550 nm. By monitoring the output port for 780 nm we measure the visible emission spectrum of the microring as shown in Fig. 4(d). We obtain strong SHG emission at the corresponding ring resonances. Limited only by the bandwidth of the EDFA we observe SHG over a bandwidth of 20.8 nm around 780 nm.

The generated second-harmonic light is strong enough to be observable with a CCD camera (Hamamatsu C8484-03G) in the free-space setup shown in Fig. 3(b). By focusing on the ring with a microscope objective we collect the light scattered out of the microring. Surface roughness leads to additional back-scattering inside the waveguide. As a result both the clock-wise (CW) and counter-clockwise (CCW) ring optical modes are excited as confirmed by the transmission spectrum of a typical resonance shown in Fig. 4(e), which exhibits a double-dip resulting from mode-splitting between the CW and CCW modes. Excitation of the counter-propagating modes inside the ring results in the emission of SHG light in both directions into the drop port. This is illustrated in Fig. 4(f) in the CCD camera image taken with longer exposure times, confirming that light from both resonating modes emits out of the ring. By analyzing the collected light with a spectrometer we verify that the wavelength of the emission corresponds to half the pump laser wavelength, with comparable magnitude on both waveguide ends. Using the free-space setup we can record the resulting standing wave pattern for the ring resonances as shown in Fig. 4(g) by consecutively tuning the pump laser wavelength to the ring resonances that fall within the bandwidth of the EDFA.

The time-independent fringe pattern is observable for SHG from 766 nm to 787 nm, demonstrating that broadband second-harmonic generation is achievable on a chip. The wavelengths corresponding to the ring resonances are ordered from left to right in ascending order.

The intensity of the emitted SHG light is convoluted by the transmission spectrum of the input grating couplers, which shows peak transmission around 1540nm. Furthermore the gain spectrum of the EDFA leads to non-uniform pump power of the pump spectrum shown in Fig.4g) and thus SHG emission from low and high wavelengths is reduced.

## 5. Power dependence and conversion efficiency of the SHG

By monitoring the power dependence of the generated SH around 780 nm we confirm the second order nature of the nonlinear process. We obtain the transmission characteristics of the device shown in Fig. 3(a) by monitoring the power coupled out of the device in dependence of wavelength. In the through port we observe the spectrum in Fig. 5(a), showing the ring resonances separated by the FSR. At high input power we can detect the cross-transmitted SH light in the output port as shown in Fig. 5(b). In Fig. 5(c) we plot the SH output power in the output waveguide in dependence of pump power on the input waveguide. We achieve a conversion efficiency of -45 dB at 120 mW pump power when 2.2 µW SH light is generated in each direction of the output waveguides. The expected quadratic dependence of the output power is clearly observed in the data, where the red line is a best fit to a quadratic dependence. By plotting the power dependence on a log-log scale we obtain a best fit to the slope of 2.03 ± 0.02, which is very close to the expected second order scaling law.

While exact phase matching can only be achieved at one wavelength, the SH intensity is generally a function of the phase mismatch $I_{sh} \sim (L/\lambda)^2 sinc^2(\Delta kL/2)$, where $L$ is the cavity length, $\Delta k$ is the wavenumber mismatch between the pump and SH. The phase matching bandwidth (in *meters*) is related to the cavity length $L$ and the group velocity mismatch by $\delta\lambda_{FWHM} = 0.44\lambda_0^2 / [Lc_0(v_{g,sh}^{-1} - v_{g,p}^{-1})]$ [25], where $v_{g,sh}$ and $v_{g,p}$ are the group velocities of the SH and the pump respectively. The relatively flat SH conversion efficiency observed in Fig. 4(b) is an indication that approximate phase matching has been achieved from 770 nm to

785 nm as a result of the small discrepancy between the group indices of the pump light ($n_{g,p}$=2.55±0.05) and SH ($n_{g,sh}$=2.4±0.4).

We can proceed to determine the magnitude of the $\chi^{(2)}$ taking into account the linear transmission properties of the grating couplers, the parameters of the ring resonators and the coupling characteristics to the output waveguide. Following the derivation described in [26] we solve the coupled mode equations for the electric fields on the waveguide and the ring resonator. The pump-wave at $\lambda_p$ is assumed to convert to a second-harmonic signal wave at $\lambda_s$ obeying the energy conservation requirement $2\omega_p=\omega_s$. Furthermore we assume in the following that exact momentum conservation is fulfilled through phase-matching inside the ring.

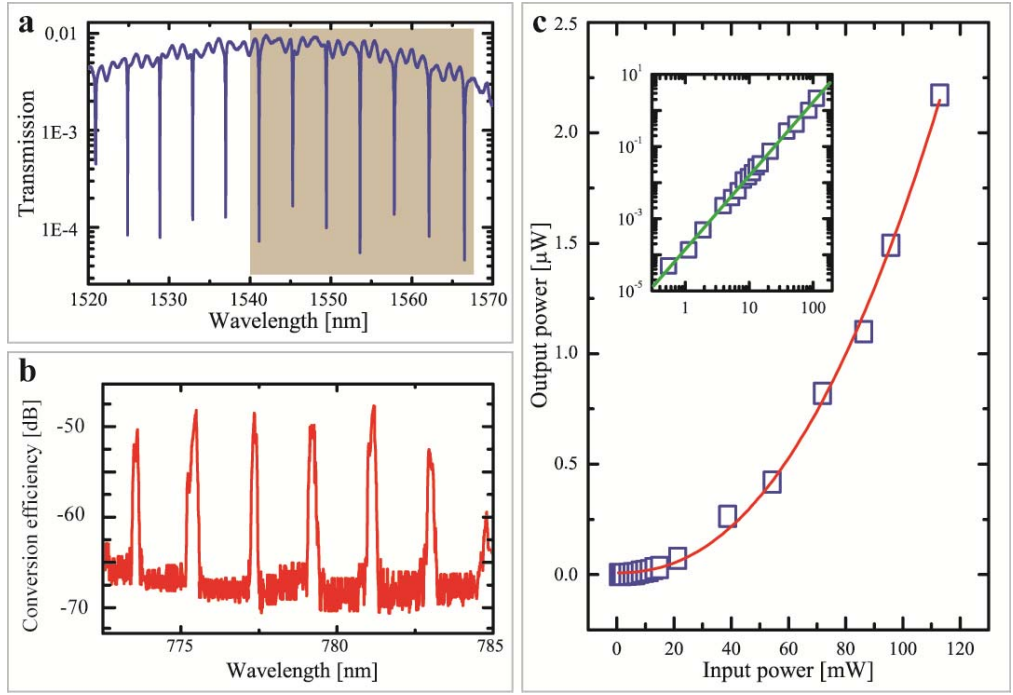

Fig. 5. Optical transmission performance and SHG power output. (a) Through port transmission spectrum of the pump light showing ring resonances separated by a FSR of 3.8 nm enveloped by the profile of the grating coupler; typical quality factors are ~10,000; the shaded grey area covers the optical resonances that are used to generated the SH light shown in panel (b). (b) the SH conversion efficiency as the pump wavelength is continuously tuned from 1540 nm to 1568 nm (grey area in panel (a)); conversion efficiency of roughly -45 dB is observed when the pump wavelength coincides with one of the ring resonances. (c) the quadratic dependence of SH power in the output waveguide on the pump power in the input waveguide (markers) and the parabolic fit (red line); Inset: the slope of the double log plot is fitted to be 2.03 ± 0.02, very close to the theoretical quadratic law.

From the electrical field amplitudes we can calculate the power of the generated signal wave and thus determine the small signal conversion efficiency as [26, 27]

$$\eta = \Gamma F_p^2 F_s \frac{L^2 \pi^2}{2 A_{eff,p}^2 \lambda_p^2} \frac{A_s}{c\varepsilon_0 n_s} \left(\chi^{(2)}\right)^2 P_p \qquad (1)$$

Here $A_s$, $A_{eff,p}$ are the effective mode areas of the SH mode and the pump mode. $n_s$ is the effective refractive indices of SH and $F_p$, $F_s$ denote the field enhancement inside the cavity for pump and SH respectively. In the above equation $L$ is the circumference of the ring resonator and $\Gamma$ denotes the field overlap factor between the second-harmonic mode $E_s$ and the pump mode $E_p$ inside the waveguide, which is given by the following equation

$$\Gamma = \frac{\left|\int E_s^* E_p \, dA_s \, dA_p\right|^2}{\int |E_p|^2 \, dA_p \int |E_s|^2 \, dA_s} \quad (2)$$

The effective mode areas and field overlap are obtained from finite-element simulations using COMSOL Multiphysics. The field enhancement factor on resonance can then be calculated [28] as

$$F = \left|\frac{\kappa}{1 - t_1 t_2 \alpha}\right| \quad (3)$$

Here $t_1$ and $t_2$ are the transmission factors on the through and the drop port respectively, $\alpha$ is the round trip loss and $\kappa$ is the coupling coefficient. From the transmission spectrum at 1560 nm and 780 nm the field enhancement factors for 1550 nm and 780 nm light are estimated to be around 24 and 8, respectively. The pump power on the waveguide is obtained from the measured input power after the fiber amplifier, taking into account the calibrated coupling loss at the grating couplers. The power of the SHG light is determined by taking into account the out-coupling loss from the 780 nm grating couplers, which was calibrated independently through transmission measurements in the drop port. Finally, we calculated the nonlinear susceptibility $\chi^{(2)}$ to be 16 ±7 pm/V, which is on the same level as the value of bulk GaN [11].

## 6. Conclusions

Due to its remarkably wide transparency window, spanning from the mid-IR all the way to visible/UV wavelengths (365 nm~13.6 μm) [12], GaN has enormous potential for linear and nonlinear optical applications. SHG in integrated GaN photonic circuits provides an efficient way to exploit these wavelength regimes which are currently unavailable to silicon photonic devices. Integration of GaN on a silicon platform has high potential to improve the performance of silicon photonic circuits by adding new functionalities, enabling better high power and high frequency electronics as well as opening new opportunities in the field of optoelectronics. The combination of GaN electronic components, light emitting and laser diodes [16] with integrated photonics hold promise for realizing a full suite of opto-electronic circuits on the same chip.

In summary, we have developed a novel photonic circuit concept based upon heterogeneous integration of GaN on silicon substrates. Efficient second harmonic generation is demonstrated in this hybrid photonic architecture. We anticipate that the $\chi^{(2)}$ nonlinearity demonstrated here will find use in the efforts to expand the bandwidth of communication, to realize non-classical light sources [2, 29] and to achieve many other crucial functions employed for on-chip nonlinear and quantum optics.

At the time of this submission, related work on the observation of second harmonic generation in an ultrahigh quality factor silicon nitride ring rsonator is reported [30]. In this case the SHG is attributed to the broken symmetry at the silicon dioxide/silicon nitride interface.

**Acknowledgements**

This work was supported by NSF, Packard Foundation and a seedling grant from DARPA/MTO (Award No.W911NF-09-1-410). W.H.P. Pernice would like to thank the Alexander-von-Humboldt foundation for providing a postdoctoral fellowship. T. Palacios and K. Ryu acknowledge support from DARPA GaN-Si integration program. The authors wish to thank Dr. Michael Rooks, Michael Power, James Agresta and Christopher Tillinghast for assistance in device fabrication.